\documentclass[journal,comsoc]{IEEEtran}
\usepackage[T1]{fontenc} 
\usepackage[utf8]{inputenc}
\usepackage[english]{babel} 
\usepackage{amsmath,amsfonts, amssymb} 
\usepackage{mathtools}
\usepackage{acronym}
\usepackage{hyperref}
\usepackage{longtable}
\usepackage{textcomp}
\usepackage{multirow}
\usepackage{makecell}
\usepackage{subfigure}
\usepackage{graphicx}
\usepackage{acronym}
\usepackage{color}
\usepackage{epsfig}
\usepackage{url}
\usepackage{balance}
\usepackage{float}
\usepackage{array}
\usepackage{comment}
\usepackage{pifont}
\usepackage{verbatim}
\usepackage{pifont}
\usepackage{tabularx}
\usepackage{hyperref}

\newcommand{\cmark}{\ding{51}}
\newcommand{\xmark}{\ding{55}}
\newcolumntype{P}[1]{>{\centering\arraybackslash}p{#1}}
\newcommand\blfootnote[1]{%
  \begingroup
  \renewcommand\thefootnote{}\footnote{#1}%
  \addtocounter{footnote}{-1}%
  \endgroup
}

\acrodef{IMO}{International Maritime Organization}
\acrodef{RF}{Radio Frequency}
\acrodef{IALA}{International Association of Lighthouse Authorities}
\acrodef{AIS}{Automatic Identification System}
\acrodef{VHF}{Very High Frequency}
\acrodef{GMSK}{Gaussian filtered Minimum Shift Keying}
\acrodef{VTS}{Vessel Traffic Service}
\acrodef{MASS}{Maritime Autonomous Surface Ships}
\acrodef{MMSI}{Mobile Maritime Service Identity}
\acrodef{HF}{High Frequency}
\acrodef{UHF}{Ultra High Frequency}
\acrodef{SATCOM}{Satellite Communications}
\acrodef{GMDSS}{Global Maritime Distress and Safety System}
\acrodef{LEO}{Low-Earth Orbit}
\acrodef{GEO}{Geostationary Earth Orbit}
\acrodef{VSAT}{Very Small Aperture Terminal}
\acrodef{NOC}{Network Operation Center}
\acrodef{DVB-S}{Digital Video Broadcasting - Satellite}
\acrodef{SSAS}{Ship Security Alarm System}
\acrodef{SOLAS}{Safety of Life at Sea}
\acrodef{GNSS}{Global Navigation Satellite System}
\acrodef{TDMA}{Time Division Multiple Access}
\acrodef{DSC}{Digital Selective Calling}
\acrodef{GLONASS}{GLObal NAvigation Satellite System}
\acrodef{BDS}{BeiDou Navigation Satellite System}
\acrodef{DoS}{Denial of Service}
\acrodef{ECDIS}{Electronic Chart Display and Information Systems}
\acrodef{BNWAS}{Bridge Navigation Watch Alarm System}
\acrodef{LEO}{Low Earth Orbit}
\acrodef{MEO}{Medium Earth Orbit}
\acrodef{GNSS}{Global Navigation Satellites Systems}
\acrodef{GPS}{Global Positioning System}
\acrodef{GLONASS}{Global Navigation Satellite System}
\acrodef{SDR}{Software Defined Radio}
\acrodef{ToA}{Time of Arrival}
\acrodef{EDR}{Event Data Recorder}
\acrodef{NMEA}{National Marine Electronics Association}
\acrodef{CI}{Critical Infrastructure}

\graphicspath{{figs/}}

\title{\textcolor{black}{Vessels Cybersecurity: \\ Issues,  Challenges, and the Road Ahead}}
\author{
    \IEEEauthorblockN{Maurantonio Caprolu, Roberto Di Pietro, Simone Raponi, Savio Sciancalepore, Pietro Tedeschi} \\
    \IEEEauthorblockA{Division of Information and Computing Technology \protect\\ College of Science and Engineering, Hamad Bin Khalifa University - Doha, Qatar
    \\\{ssciancalepore, rdipietro\}@hbku.edu.qa, \{mcaprolu, ptedeschi, sraponi\}@mail.hbku.edu.qa}
}

\begin{document}

\maketitle

\begin{abstract}
\blfootnote{This article has been accepted by IEEE Communications Magazine. Publication expected in June 2020.}
Vessels cybersecurity is recently gaining momentum, as a result of a few recent attacks to vessels at sea. These recent attacks have shacked the maritime domain, which was thought to be relatively immune to cyber threats. The cited belief is now over, as proved by recent mandates issued by the International Maritime Organization (IMO). According to these regulations, all vessels should be the subject of a cybersecurity risk analysis, and technical controls should be adopted to mitigate the resulting risks.
This initiative is laudable since, despite the recent incidents, the vulnerabilities and threats affecting modern vessels are still unclear to operating entities, leaving the potential for dreadful consequences of further attacks just a matter of ``when'', not ``if''.

In this contribution, we investigate and systematize the \mbox{major} security weaknesses affecting systems and communication technologies adopted in modern vessels. Specifically, we describe the architecture and main features of the different systems, pointing out their main security issues, and specifying how they were exploited by attackers to cause service disruption and relevant financial losses.
We also identify a few countermeasures to the introduced attacks.
Finally, we highlight a few research challenges to be addressed by industry and academia to strengthen vessels security.

\end{abstract}

\section{Introduction}
\label{sec:intro}

Vessels are likely the oldest long-range transportation means used by humans to reach physically far locations, and it is still the preferred one in many cases, including the movement of goods---over 90\% of the world's trade is carried by sea---and luxury entertainment, such as cruise experiences. 
Thus, large vessels carrying thousands of tonnes of goods (e.g. oil tankers, containers carriers) or a few thousand people definitively can be considered as \emph{critical systems}, requiring reliable and secure computing and communication systems.

However, vessels cybersecurity issues historically have received only minimal attention from both the shipowners and the scientific community. The reasons are manifold, and can be found in the late \emph{digitization} of the maritime sector, the heterogeneity of operators, the focus on availability rather than security, and, finally, the widespread belief that cyber-attacks to vessels offshore are technically difficult and hard to succeed.


\textcolor{black}{Given the above scenario, the \ac{IMO} first generally highlighted the major cyber-threats in the maritime sector (\emph{MSC-FAL.1 /Circ.3}, 2017). Then, in the 98th session (June 2017), it took action, adopting the ``Resolution MSC.428/98'', namely, ``Maritime Cyber Risk Management in Safety Management Systems''. This document forces shipowners to address cyber risks and cyber-security attacks in the design and the deployment of existing safety management systems~\cite{imo2019}---these objectives to be attained by January 2021. 
Since this latter date, any vessel not conforming to this regulation would not be authorized to sail---implying relevant economic loss, as well as a threat to global trade.}

Despite this mandate, the \ac{IMO} did not clearly define the attack surface of a modern vessel. Indeed, while shipowners are currently oriented towards the strengthening of the management systems, other areas are overlooked. For instance, severe weaknesses implicit to the communication protocols used by the ships are unknown to operating entities, hence potentially nullifying the efforts to implement secure-by-design vessels. 

\textcolor{black}{To the best of our knowledge, this is the first contribution investigating, in a comprehensive and structured manner, the cybersecurity issues associated with modern vessel systems. Specifically, we analyze the communication technologies and computer systems used within large vessels, pointing out several security issues rooted in their design and operational mode. We also relate these vulnerabilities with recent incidents and attacks involving vessels, and we identify the weak points that any modern vessel needs to mitigate towards the enforcement of the latest IMO resolutions. Finally, we indicate   mitigating countermeasures to the highlighted threats, as well as  future research directions to be addressed by industry and academia to strengthen vessels cybersecurity}. 

The rest of this paper is organized as follows: Section~\ref{sec:comm} discusses the security vulnerabilities of communication technologies on vessels, 
Section~\ref{sec:vessel} illustrates the vulnerabilities associated with the computer systems on-board, while Section~\ref{sec:safety} focuses on the technologies used to improve people' safety. Section~\ref{sec:discussion} highlights challenges, countermeasures, and future directions towards the realization of cyber-secure vessels. Finally, Section~\ref{sec:conclusion} tightens conclusions.

\begin{figure*}[htbp]
  \centering
  \includegraphics[angle=0, width=\textwidth]{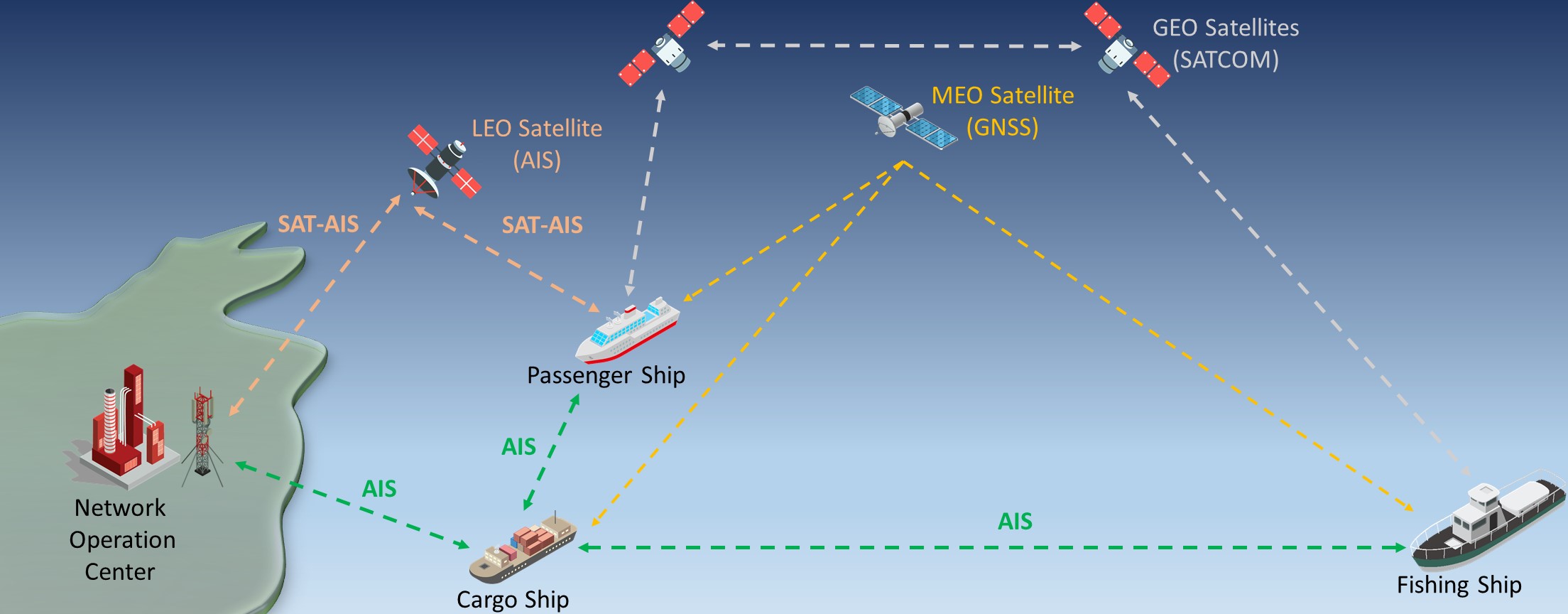} 
  \caption{Overview of technologies used by modern vessels.}
  \label{fig:scenario}
\end{figure*}

\section{Vessel Communication Systems Security}
\label{sec:comm}
Traditional inland communication systems have not been designed to guarantee network coverage offshore. Thus, only a limited set of technologies are available for vessel communications.
Figure~\ref{fig:scenario} provides a graphical overview.
While GNSS technologies enable location estimation (Section~\ref{sec:gnss}), critical bidirectional communication services include the \ac{AIS} protocol (Section~\ref{sec:ais}) 
and \ac{SATCOM}---this one supporting higher data rates (Section~\ref{sec:sat}). 

\subsection{Global Navigation Satellites Systems (GNSS)}
\label{sec:gnss}

A key element of modern vessel systems is the real-time location awareness provided by \ac{GNSS} technologies. Any \mbox{modern} vessel is equipped with a GNSS module, that can receive RF signals originated from \ac{MEO} satellites, located $\left[19,000 - 23,000\right]$~km above Earth. GNSS data, guaranteeing earth coverage, are broadcasted unidirectionally at a frequency of $2$~Hz.
Each satellite is synchronized to the exact system time, thanks to atomic clocks, and transmits a navigation signal containing the message delivery time and additional information, including the deviation of the satellite from its expected trajectory. 

A receiver equipped with omnidirectional antennas detects a combination of signals from different satellites and can identify the single contributions. For each of them, based on the exact \ac{ToA} (synchronized with the clock reference of the transmitters), knowing the propagation speed of the signal (the speed of light), it is possible to obtain the distance from the satellites. A minimum of four satellites are required to efficiently multi-laterate distances and obtain a location. Given that perfect time synchronization is not possible, a localization error is generally present, usually not exceeding $5$~meters in outdoor conditions with clear sky visibility~\cite{oligeri2019_wisec}. Note that, when employed for military use (access is restricted to authorized parties), the GNSS can reduce the error to less than 1 meter.
Several \ac{GNSS} technologies are available, based on the community responsible for operating and maintaining the satellites. Despite the most famous is the \ac{GPS} operated by the USA, there are equivalent systems, e.g., the Russian \ac{GLONASS}, the European GALILEO, and the Chinese BEIDOU.

\begin{figure}[htbp]
  \centering
  \includegraphics[angle=0, width=.9\columnwidth]{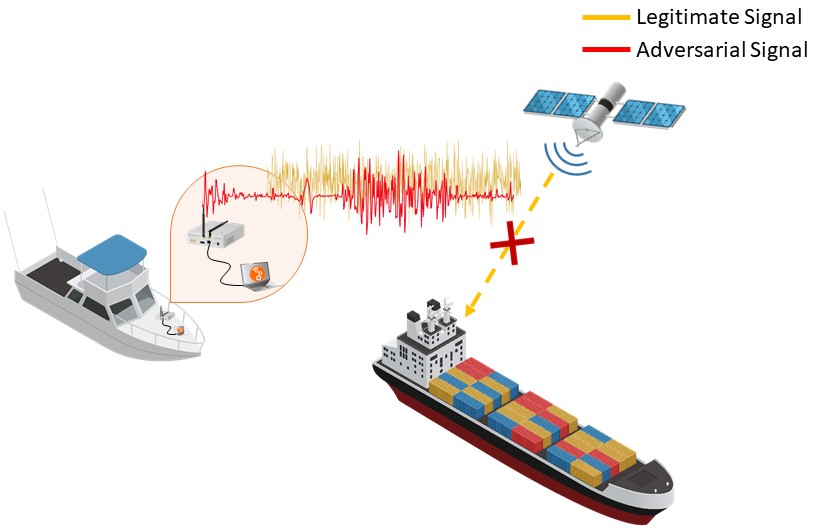}
  \caption{Logic of jamming and spoofing attacks against vessels.}
  \label{fig:jamming}
\end{figure}

From the security perspective, commercial vessels rely on civilian \ac{GNSS} signals. Unfortunately, to boost message availability at the receivers, the civilian \ac{GNSS} was designed to transmit messages in clear-text, without relying on any confidentiality nor authentication mechanism. Moreover, since \ac{GNSS} signals are used as the timing source of many synchronization technologies, the trajectories of the satellites are publicly available.
Therefore, they can be easily spoofed using commercially available \acp{SDR}. These relatively cheap devices can be tuned on the operating frequency of the \ac{GNSS} technology, and configured via freely-available, open-source tools to transmit messages that are indistinguishable from authentic \ac{GNSS} signals. As shown in Figure~\ref{fig:jamming}, since the legitimate signals are very weak at the ground level, the forged messages can be easily super-imposed to the legitimate ones, leading the receivers to estimate fake locations~\cite{baldini2011_comst}. 

Despite these weaknesses are well-known, the vulnerability of vessels to \ac{GNSS} spoofing attacks received worldwide attention only recently, due to anomalies detected in the Black Sea area \cite{blacksea2017}. 
In June 2017, at least 20 ships located in the Black Sea region reported anomalies to their \ac{GNSS} systems, with their receivers signaling either unstable locations or positions on the ground, close to a nearby airport area. Thus, vessels were forced to switch to manual navigation systems and to resort to outdated systems to maintain their routes and avoid collisions.
\ac{GNSS} is extremely sensitive also to jamming attacks.
Vessels can be easily approached by moving entities carrying a device emitting noise on the \ac{GNSS} communication frequency. The power of the noise adds up to the power of the legitimate signal, thus compromising the operation of any \ac{GNSS} technology.

\subsection{Automatic Identification System (AIS)}
\label{sec:ais}
\ac{AIS}, proposed by the \ac{IALA} and part of the \ac{VTS}, is a coastal tracking system mandatory on ships of over 300 tonnes, but widely adopted also over ships of smaller weight. 
It aims to broadcast the position, speed, movements, and route of the vessels, to help them avoiding collisions~\cite{Tu2018}. \ac{AIS} operates in the \ac{VHF} band, using the \ac{GMSK} scheme, and it guarantees a horizontal transmission range of up to $74$~Km, 
with a bitrate of $9600$~bit/s. \ac{AIS} uses two channels: the $161.975$~MHz, for ship-to-ship communication, and the $162.025$~MHz, for ship-to-shore communications. \ac{AIS} transceivers consist of: (i) a \ac{VHF} transmitter; (ii) two \ac{VHF} \ac{TDMA} receivers; (iii) a \ac{VHF} \ac{DSC} receiver; (iv) a positioning module supporting \ac{GNSS} capabilities; and (v) sensors connected via standard marine electronic communications links. The data are periodically broadcast, allowing all the compatible vessels to enforce situation awareness and avoid collisions.

To overcome coverage issues, \ac{AIS} has been extended further to a space-based version, namely Space-Based \ac{AIS}. Leveraging \ac{AIS} receivers located on \ac{LEO} satellites, the messages can be relayed to the ground, extending the range up to $400$~Km. 

While the intended purpose of AIS was to avoid vessel collisions, today \ac{AIS} is used for several cyber-physical applications, including identification, search and rescue operations, accident investigation, remote tracking, ocean currents estimation, and the protection of marine \acp{CI}.

However, being designed in the 80s, \ac{AIS} does not support any security property, such as authentication and confidentiality. As discussed in~\cite{Balduzzi2014}, the protocol is vulnerable to different attacks, including spoofing, hijacking, data manipulation, and \ac{DoS}. An attacker could: (i) create fake vessels; (ii) inject false ship details (e.g., position, speed, 
and \ac{MMSI}); (iii) impersonate vessels or port authorities; (iv) inject false 
information (e.g., false man-in-water alarms); 
and, finally, (v) send false collision warning alerts.

\subsection{Satellite Communications}
\label{sec:sat}

\ac{SATCOM} services are the roots of many services used on vessels to guarantee safety and security.
A generic SATCOM system consists of four elements: (i) several space-based satellites; (ii) many ground-based gateway earth stations; (iii) a \ac{VSAT} antenna installed outdoor; and, (iv) a \ac{NOC}. 

The satellites are the key elements of the networking infrastructure. They are typically located in a \ac{GEO} (height $35,786$~Km) above the equator, and they are equipped with multiple communication technologies to communicate with earth-based equipment and other satellites. Typically, a GEO satellite exchanges data using the Ka-band, in the frequency range $\left[26.5 - 40\right]$~GHz, using narrow-band modulation schemes. This allows re-using the frequency band, compared to traditional wideband technologies. Alternatively, the \emph{bent-pipe} architecture is used, where the satellites act as relays in \emph{Amplify-and-Forward} mode, relaying the signal between two gateways located on the ground. 
The satellites can also use dedicated optic technologies to transfer data to each other, as for \ac{LEO} satellites operated by GlobalStar and Iridium providers.

In maritime applications, the gateway earth stations and the VSAT antenna reside together on the vessel, and enable direct communication with \ac{GEO} satellites at high speed, up to 506 Mb/s when using the most recent Ku-band (frequency range: $\left[12 - 18\right]$~GHz). 
To guarantee Line-Of-Sight (LOS) connection to the satellites, the VSAT antenna is installed outdoor, in a clear view of the sky with a given angular view, expressed in terms of azimuth, polarization, and skew. This setup is typically realized by the operator at the deployment time. Motors and sensors on-board are used to orientate the antenna towards the satellite. 
Finally, the NOC is maintained by the satellite operator and provides coordination and maintenance tasks to the satellite network. Thus, each ship is typically a node in a network including also vessels of the same shipowner and same satellite operator. Many maritime satellite providers are available, including Iridium, GlobalSat, and INMARSAT. INMARSAT, operating 13 satellites, is the most adopted operator, as it is the only approved provider for the \ac{GMDSS} technology.

The \ac{GMDSS} specification, established in 1980 by the \ac{IMO}, 
evolved to include several systems, protocols, equipment, and procedures useful to ease the rescue of vessels in distress. It includes: (i) transmission of ship-to-shore distress communications via multiple communication technologies on multiple frequencies; (ii) reception of shore-to-ship alerts; 
(iii) transmission of maritime safety information; (iv) transmission of vessel location; and, (v) exchange of generic navigation information. 

Overall, the degree of security offered by SATCOM strongly depends on the specific protocols and operations deployed by the operator. Since operators are private companies, the details of the security protocols and procedures are always protected by intellectual property rights, thus being hard to evaluate.
To name a few cases of vendor-specific security issues, a recent study~\cite{Pavur2019_wisec} discovered unencrypted connections between the VSAT antenna and the gateway in DVB-S SATCOM networks. Thus, when insecure services are used (i.e., POP3 e-mails or HTTP browsing), privacy issues arise. 

In the specific context of vessels, recent reports published by the security firm IOAlliance discovered that the GMDSS operations enforced by INMARSAT are affected by severe weaknesses~\cite{IOActive}. The root cause of the vulnerabilities has been found in the \emph{thraneLink} protocol, a proprietary solution within the SAILOR 6000 communication suite. Specifically, the consulting firm found a backdoor enabling the installation of an unauthenticated firmware update and malicious software. 
Moreover, additional protocol-level vulnerabilities were discovered in the Mini-C INMARSAT Terminal. Using specially crafted messages, an attacker aware of these vulnerabilities could disable the \ac{SSAS}, used by vessels to signal piracy and terrorism attempts offshore. Besides, additional physical attacks are possible, by simply modifying the orientation of the VSAT antenna, thus denying a reliable SATCOM link.

\section{Vessels Computer Systems Security}
\label{sec:vessel}

Computer systems integrated into modern vessels include specific hardware and software solutions to automate dedicated functions, including navigation, propulsion, and fuel supply. On the one hand, they provide the crew with a real-time and reliable view of the state of the vessel, improving reaction time and decreasing personnel costs. On the other hand, these technologies increase the vessel attack surface, leading to additional security concerns.
\begin{figure*}[htbp]
  \centering
  \includegraphics[angle=0, width=1.2\columnwidth]{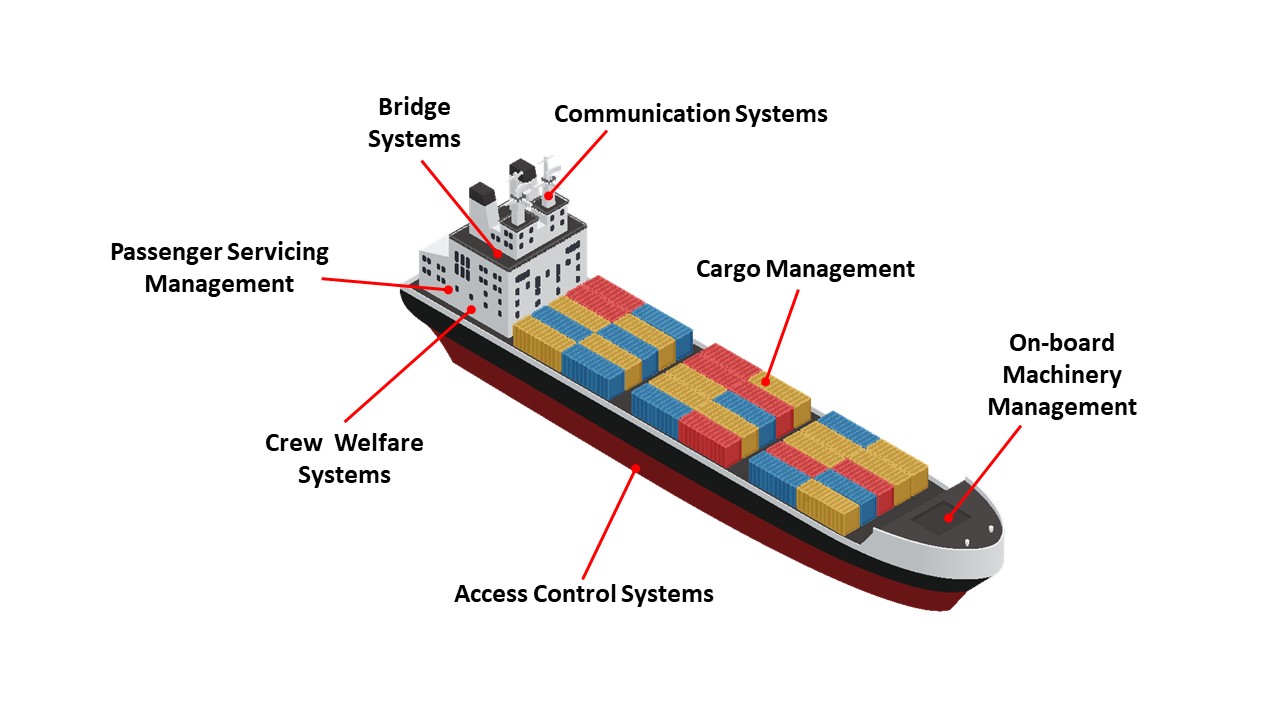}
  \caption{Main systems integrated into modern vessels.}
  \label{fig:vessel_attack_surface}
\end{figure*}

The automated systems operating on-board, summarized in Figure~\ref{fig:vessel_attack_surface}, include a variety of elements~\cite{bimco2016guidelines}. The core is the \emph{bridge}, providing a unified view of all the systems operating on-board. The bridge acts as the central point of a logical star network, where information coming from peripheral systems are integrated to provide a comprehensive view of the vessel. 
Another important system is the \ac{ECDIS}, a computer-based navigation information system used by officers to establish and maintain the route, as an alternative to legacy paper nautical charts. The ECDIS provides crucial services, including navigational safety, automatic route planning, route monitoring, navigation time, and route update management. 

Additional peripheral management systems connected to the bridge include: (i) the \emph{Cargo Management System}, in charge of managing goods loading and unloading operations; (ii) the \emph{access control systems}, including cameras and microphones for surveillance purposes, alarm systems, and electronic devices for the on-board personnel security; (iii) \emph{on-board machinery management}, automating the monitoring of mechanical systems, including propulsion and steering systems; and, finally, (iv)  the \emph{communication systems}, including SATCOM, AIS, and GNSS modules.

Moreover, the ECDIS provides continuous data recording features, thanks to the interaction with the \ac{EDR}. The \ac{EDR} is an event logger allowing forensic investigations when serious malfunctions and incidents occur. According to \ac{SOLAS} regulations, the EDR should provide minute-by-minute recording for the past twelve hours of the voyage and the record of four hourly intervals of voyage track for a period of six months. These systems could be connected to shore-side networks for data downloading and software updates.

Other computer systems are available, even if they are mostly physically disconnected from the bridge. To name a few, they include the \emph{crew welfare systems}, integrating several third-party computing systems used for ship administration and crew welfare, and the \emph{passenger services management}, including the devices in possession by the crew and the passengers and used for boarding, access control, billing services, luggage tracker, and entertainment.

\textcolor{black}{From the security perspective, being vessels Cyber-Physical Systems, the physical and the digital components of the system are interrelated. Thus, attacks on the digital infrastructure impact on the physical context of the vessel, and, at the same time, physical attacks can disrupt the digitized systems maneuvering and coordinating vessel operations.}
In particular, the \emph{bridge} is the most critical component. Gaining full access to this system would enable the attacker to definitively control the vessel, performing maneuvers, and altering the input from peripheral systems. 
Severe service disruption could occur also if the attacker takes full control of crucial peripheral systems, including the EDR, cargo management systems, and the on-board machinery management systems. 
\textcolor{black}{For instance, a ransomware controlling the vessel could block any door or movement toward the land, holding passengers as hostages at sea until a ransom is paid. On the physical perspective, an attacker could tamper the sensors of the automatic docking system, leading the automatic navigation systems to run the ship into natural formations (e.g., underwater rocks), or human infrastructures like bridges, ports, and other ships, using the very same mass and speed of the vessel as a weapon~\cite{jones2016threats}.}

News about attacks on vessel computer systems are widely available. For instance, the ``Guidelines on Cyber Security on-board Ships'' reported that a vessel designed for paperless navigation was delayed from sailing for days after a malware blocked its \ac{ECDIS} system~\cite{bimco2016guidelines}. The crew members did not realize the failure as a cyberattack, but simply as a technical issue, and its resolution took significant time and efforts, leading to relevant financial losses. 
Several shipowners also reported that their ICT network was infected by ransomware, causing service breakdowns~\cite{lee2018}. Despite the involved companies revealed the least possible details (a disappointingly diffused habit, theoretically reducing bad publicity, but practically preventing to assess the scale and impact of the phenomenon), unofficial news reported unwitting ship agents as the source of the malware, causing issues in several ports.

\section{Safety On-board}
\label{sec:safety}

Vessels are technically sophisticated systems, characterized by a considerable size. Thus, they need several crew members on-board, holding critical roles to safeguard the passengers, the vessel, and the goods. 

Several use-cases are possible for the above scenario. For instance, during the ship's docking maneuver, failing of the automatic systems could occur. In these cases, specialized crew members are typically in charge of solving the problem. However, if these people have any issue, the vessel does not have any technology enabling a timely communication of the event or the automatic handling of the emergency.

Similarly, the unexpected absence of any crew member during critical real-time vessel operations could not be automatically detected. At this time, the control room where the bridge is located is mostly unable to realize personnel absence and undertake a critical maneuver without being supported by specialists.

To the best of our knowledge, the safety on-board is currently addressed by relying on both introductory training courses and legacy communication protocols. 
During the training courses, the crew members acquire general information, such as reaction behaviors in case of fire, safety equipment usage, and alarm signals meaning. However, no information about the operation of communication systems is provided.

As per the communication protocols, the \ac{NMEA}-2000 is the wired communication standard adopted by vessels. It is a plug-and-play technology (IEC 61162-1), allowing the connection of marine sensors within vessels and their management through the bridge. The standard is compatible with the Controlled Area Network (CAN-BUS), currently employed in many vessels to manage several peripheral systems, including the steering and the alarm systems.

\begin{figure}[htbp]
  \centering
  \includegraphics[angle=0, width=\columnwidth]{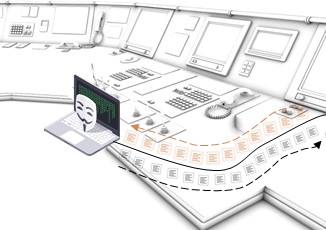}
  \caption{CAN-BUS attacks. Compatible receivers connected to the system can read and inject fake messages.}
  \label{fig:canbus}
\end{figure}

The implementation of a standard communication protocol on-board allows automating many of the tasks that, until a few decades ago, were purely manual. However, the CAN-BUS was designed in the 80s; thus, the technology proves not to be up to today's cyber-security challenges. Among the many limitations, the protocol provides neither authentication nor confidentiality of messages. Thus, as shown in Figure~\ref{fig:canbus}, any compatible receiver attached to the system could read the unencrypted content of the messages and inject fake messages---e.g., shutting down the systems, causing \ac{DoS} attacks~\cite{carsten2015vehicle}.

\section{Challenges and Road Ahead}
\label{sec:discussion}

\textcolor{black}{Table~\ref{tab:security} summarizes the security properties enjoyed (or missing) by the communication \mbox{technologies} used on vessels.}
\begin{table}[htbp]
\caption{Security properties of vessels communication technologies.}
\centering
    \begin{tabular}{|c|c|c|c|}
    \hline
        \textbf{Technology} & \textbf{Confidentiality} & \textbf{Authentication} & \textbf{Availability} \\
        \hline
        GNSS & \xmark & \xmark & \xmark \\
        \hline
        AIS & \xmark & \xmark & \xmark \\
        \hline
        SATCOM & \xmark & \cmark & \cmark \\
        \hline
        CAN-BUS & \xmark & \xmark & \cmark \\
        \hline
    \end{tabular}
\label{tab:security}
\end{table}
Despite the vulnerabilities highlighted in the above table have been addressed in other contexts, their integration in modern vessels leads to several challenges (summarized in Table~\ref{tab:challenges}), due to the constraints peculiar to the vessel context. Detailed motivations, challenges, and future directions stemming from discussions with major players in the vessels production domain are provided below.
\begin{table}[htbp]
\caption{Overview of security issues and possible countermeasures.}
\centering
    \begin{tabular}{|P{3.2cm}| P{4.8cm}|}
    \hline
        \textbf{Security Issue} & \textbf{Countermeasure} \\
        \hline
        GNSS Spoofing & Cross-Technology Location Estimation (GNSS, SATCOM) \\ \hline
        Electronic Warfare & Anti-jamming Protocols \\ \hline
        Non-Standardized SATCOM Protocols & Standardization Efforts \\ \hline
        AIS Spoofing & Software Security Frameworks \\ \hline
        Bridge System Assessment & Standardized Security Assessment Procedures \\ \hline
        Malware Attacks & Containerization \\ \hline
        Automatic Safety Systems & Wireless Sensing and ML \\ \hline
        Wired Communication Protocols Security & Physical Security Strategies, Access Control \\
        \hline
    \end{tabular}
\label{tab:challenges}
\end{table}

\textcolor{black}{
\textbf{\ac{GNSS} Spoofing Detection.} The lack of message authentication in civilian \ac{GNSS} technologies exposes vessels to \ac{GNSS} spoofing attacks, leading them to estimate inconsistent locations. Several strategies are available to detect \ac{GNSS} spoofing attacks. 
These techniques rely either on multiple antenna schemes or on the analysis of the \emph{raw} \ac{GNSS} signals, or on the cross-validation of \ac{GNSS}-derived locations with information from additional communication technologies (i.e., cellular networks)~\cite{oligeri2019_wisec}.
However, these schemes are hardly applicable to vessels. On the one hand, they require hardware modifications on systems already deployed, leading to consistent hardware installation and maintenance costs. On the other hand, vessels located off-shore could hardly leverage positioning technologies other than the \ac{GNSS}.  
Therefore, future GNSS Spoofing detection schemes on vessels should target non-invasive solutions, requiring little (if any) hardware change to the already operational expensive ships. \\ \\
\textbf{Electronic Warfare Mitigation.} Electronic Warfare scenarios involve several powerful attacks, where adversaries use sophisticated tools to disrupt the operation of the communication infrastructure of a given entity. 
In the maritime domain, jamming vessels could have disastrous consequences. Indeed, without access to the \ac{GNSS} infrastructure and to the SATCOM network, a vessel could easily lose situational awareness and the capability to communicate with landline systems, not relying on any further help than its personnel and past generations equipment (e.g., physical maps). 
Despite the availability of several anti-jamming schemes, vessels require the adoption of non-invasive protocols, that should be thoroughly assessed and contextualized 
in order not to require expensive and time-consuming hardware change operations, while providing seamless integration with current technologies and ease of use.\\\\
\textbf{SATCOM Security Standardization.} The recent attacks over SATCOM networks have revived the interest of industries and academia towards the security of the satellite communication links. To date, the high costs derived from the manufacturing, deployment, and maintenance of satellites have motivated the development of proprietary solutions. Thus, the involved protocols are (despicably) mostly closed-source and undocumented, and their security cannot be evaluated from the scientific community---it may be worth remembering that security through obscurity, while providing a transient competitive advantage, in the long-run has been shown to lead to critical breaches. Indeed, these limitations indicate the need for standardization activities towards the definition of secure datalinks for vessels SATCOM security. While programs specific for other CIs are starting to arise (see~\cite{bernsmed2017_dasc} for civil aviation), to the best of our knowledge, no initiatives are scheduled for the maritime domain. \\\\
\textbf{AIS Spoofing Detection.} The \ac{AIS} protocol provides neither 
message authentication nor encryption, thus being exposed to replay and spoofing attacks. Similar issues existing for other wireless communication technologies have been addressed thanks to application-layer frameworks, able to provide authenticity, integrity, and confidentiality to the messages. These cited strategies can be integrated into the  applications using AIS, leveraging the lessons learned by other similar communication technologies (see IEEE 802.15.4 in the IoT context). However, conjoint initiatives by shipowners and operators in the vessel domain are needed to agree on application-layer design guidelines and technical details, such as the setup of a dedicated Public Key Infrastructure. \\\\ 
\textbf{Bridge Systems Assessment.} Considering the critical role played by the bridge and the catastrophic consequences that could arise in case of attacks, protecting the bridge software against cyber-attacks must be a priority for shipowners. Research efforts are needed in this field to standardize the security assessment procedures, inheriting experience from other CI domains (e.g., smart grids, aircraft), and contextualizing the maritime requirements in  security frameworks characterized by a high level of automation. \\\\
\textbf{Mitigating Malware Attacks.} Similarly to other digitized CIs, malware infections to vessel management and computing systems could have a high pay-off for an attacker. Protecting the security perimeter of a CI has been extensively studied in the literature~\cite{Puthal2017}, and similar solutions could be integrated into vessel systems as well. Given that the effective physical separation and logical isolation among all the on-board systems are imperative to contain the spread of potential threats and the extent of damages in case of malware attacks,  global regulations and technical guidelines specifically tailored to vessel systems are needed, to standardize interfaces and interconnections among integrated systems.\\\\
\textbf{Automatic Safety Systems.} Although technological innovations proceed at a high pace, their integration into modern vessels is still in its infancy. Unlike other moving assets (e.g., aircraft), vessels still largely rely on human intervention to manage emergencies. This leads to the necessity of introducing automatic emergency systems, allowing responsive monitoring of goods and humans onboard.
Wireless sensor-based systems relying on wearable embedded devices could provide a valuable solution to the above issues. Thanks to short-range wireless communication technologies, smart sensors can monitor the physical conditions of the crew, 
enabling real-time personnel localization and tracking. Moreover, optimized deployment strategies can improve situational awareness during critical maneuvers. 
For instance, a central computing system relying on artificial intelligence could collect and analyze the measurement, react to emergencies, and provide significant advantages in terms of prevention, detection, and management.  \\\\
\textbf{Wired Communication Protocols Security.} The rapid obsolescence of the current technologies makes protocols introduced decades ago no longer fit to mitigate modern cyber threats. Specifically, being proprietary protocols closed-source and protected by intellectual property rights (e.g., the CAN-BUS), their security evaluations is difficult, to say the least. Hence, there could be hidden security threats getting along for decades, as it has been recently shown~\cite{lipp2018}. This scenario could lead to perilous use-cases where, besides cargo and physical resources, even the safety of the crew members and passengers is at stake.
Thus, securing the interactions between the centralized computing equipment and the mechanical peripheral devices, as well as employing standardized and secure (ideally wired) communication protocols, become crucial security requirements, where their deployment could rely on the results already available in the scientific literature.}






\section{Conclusion}
\label{sec:conclusion}

The mandatory adoption of cyber-security risk analysis and related technical controls on all vessels by January 2021 forces shipowners to start reflecting on a few key security elements, including the threat model affecting the maritime domain, the attack surface, and possible countermeasures. In this paper, to the best of our knowledge, we highlight for the first time the main weaknesses affecting the communication technologies and systems used in modern vessels, shading lights on their relationship with mainstream attacks carried out over the last few years. We also summarize the most important research challenges, directions, and countermeasures, to be addressed by maritime operators towards the development of secure vessel systems. 

Despite the widely known security issues of wireless and wired technologies are even amplified in the vessel domain, we believe that vessels integrate a suitable variety of technologies and the computational power to overcome such limitations, meeting standard, and evolving towards more secure operational conditions. 
Though, this can only be achieved via a stricter collaboration among  maritime sector, industry, and academia.

\section*{Acknowledgments}
The  authors  would  like  to  thank  the  anonymous  reviewers for their comments and suggestions, that helped improving the quality of the manuscript.
This publication was partially supported by awards NPRP 11S-0109-180242,  UREP 23-065-1-014,  and  
NPRP  X-063-1-014 from the QNRF-Qatar National Research Fund, a member of The  Qatar  Foundation.  The  information  and  views  set  out  in this publication are those of the authors and do not necessarily reflect the official opinion of the QNRF.

\bibliographystyle{IEEEtran}
\bibliography{references}

\section*{Biographies}
\noindent {- Maurantonio Caprolu} is PhD student at HBKU-CSE-ICT, Doha-Qatar. He received his Master's Degree with honors in Computer Science at Sapienza, University of Rome, Italy. His research interests include security issues in Blockchain-based systems, Edge/Fog architecture and Software-Defined-Networking. \\
- Dr. {Roberto Di Pietro},  ACM Distinguished Scientist, is full professor of cybersecurity at HBKU-CSE, Doha-Qatar. 
 His research interests include Distributed Systems Security, Wireless Security, OSN Security, and Intrusion Detection, leading to $220+$ scientific publications and patents. 
 As for Google Scholar, he has been totaling 8400$+$ citations, with \mbox{h-index=44}, and \mbox{i-index}=120. \\
- {Simone Raponi} is PhD Student at HBKU-CSE-ICT, Doha-Qatar. He received both his Bachelor and Master Degree with honors in Computer Science at Sapienza University of Rome, Italy.
His research interests include cybersecurity, Privacy, and Artificial Intelligence.\\
- Dr. {Savio Sciancalepore}
is Post-Doc at HBKU-CSE-ICT, Doha-Qatar. He received his bachelor and master degrees from the Politecnico di Bari, Italy. 
His research interests cover security issues in Internet of Things and Cyber-Physical Systems.\\
- {Pietro Tedeschi} is PhD Student at HBKU-CSE-ICT, Doha-Qatar. He received his Master's degree with honors in Computer Engineering at Politecnico di Bari, Italy. He worked as Security Researcher at CNIT, Italy, for the EU H2020 SymbIoTe. His security research interests lie in Drone, Wireless, IoT, Applied Cryptography. 
\end{document}